\documentclass[apj]{emulateapj}

\slugcomment{To appear in The Astrophysical Journal}

\shorttitle{Environments of AGN in zCOSMOS}
\shortauthors{Silverman et al.}

\begin{document}


\title{The Environments of Active Galactic Nuclei within the zCOSMOS
Density Field}


\author{
J.~D.~Silverman,\altaffilmark{1}
K.~ Kova\v{c}\altaffilmark{1},
C. ~Knobel\altaffilmark{1},
S.~Lilly\altaffilmark{1},
M.~Bolzonella\altaffilmark{4},
F.~Lamareille\altaffilmark{5},
V.~Mainieri\altaffilmark{3},
M.~Brusa\altaffilmark{2},
N.~Cappelluti\altaffilmark{2},
Y.~ Peng\altaffilmark{1},
G.~Hasinger\altaffilmark{2},
G.~ Zamorani\altaffilmark{4},
M.~ Scodeggio\altaffilmark{8},
T.~Contini\altaffilmark{5},
C. M.~Carollo\altaffilmark{1},
K.~Jahnke\altaffilmark{18}
J.-P.~ Kneib\altaffilmark{7},
O.~ Le Fevre\altaffilmark{7},
S.~Bardelli\altaffilmark{4},
A.~Bongiorno\altaffilmark{2},
H.~Brunner\altaffilmark{2},
K.~Caputi\altaffilmark{1},
F.~Civano\altaffilmark{16},
A.~Comastri\altaffilmark{4},
G.~ Coppa\altaffilmark{4},
O.~ Cucciati\altaffilmark{6},
S.~ de la Torre\altaffilmark{7},
L.~ de Ravel\altaffilmark{7},
M.~Elvis\altaffilmark{15},
A.~Finoguenov\altaffilmark{1},
F.~Fiore\altaffilmark{13},
P.~ Franzetti\altaffilmark{8},
B.~ Garilli\altaffilmark{8},
R.~Gilli\altaffilmark{4},
R.~Griffiths\altaffilmark{17},
A.~ Iovino\altaffilmark{6},
P.~ Kampczyk\altaffilmark{1},
J.-F.~ Le Borgne\altaffilmark{5},
V.~ Le Brun\altaffilmark{7},
C.~Maier\altaffilmark{1},
M.~ Mignoli\altaffilmark{4},
R.~ Pello\altaffilmark{5},
E.~ Perez Montero\altaffilmark{5},
E.~ Ricciardelli\altaffilmark{12},
M.~ Tanaka\altaffilmark{3},
L.~ Tasca\altaffilmark{7},
L.~ Tresse\altaffilmark{7},
D.~ Vergani\altaffilmark{4},
C.~Vignali,\altaffilmark{9},
E.~ Zucca\altaffilmark{4},
D.~ Bottini\altaffilmark{8},
A.~ Cappi\altaffilmark{4},
P.~ Cassata\altaffilmark{7},
M.~ Fumana\altaffilmark{8},
C.~ Marinoni\altaffilmark{10},
H.~ J. McCracken\altaffilmark{11},
P.~ Memeo\altaffilmark{8},
B.~ Meneux\altaffilmark{2,16},
P.~ Oesch\altaffilmark{1},
C.~ Porciani\altaffilmark{1},
M.~Salvato\altaffilmark{14}
}


\altaffiltext{1}{Institute of Astronomy, ETH Z\"urich, CH-8093, Z\"urich, Switzerland.}
\altaffiltext{2}{Max-Planck-Institut f\"ur extraterrestrische Physik, D-84571 Garching, Germany}
\altaffiltext{3}{European Southern Observatory, Karl-Schwarzschild-Strasse 2, Garching, D-85748, Germany}
\altaffiltext{4}{INAF Osservatorio Astronomico di Bologna, via Ranzani 1, I-40127, Bologna, Italy}
\altaffiltext{5}{Laboratoire d'Astrophysique de Toulouse-Tarbes, Universit\'e de Toulouse, CNRS, 14 avenue Edouard Belin, F-31400 Toulouse, France}
\altaffiltext{6}{INAF Osservatorio Astronomico di Brera, Milan, Italy}
\altaffiltext{7}{Laboratoire d'Astrophysique de Marseille, Marseille, France}
\altaffiltext{8}{INAF - IASF Milano, Milan, Italy}
\altaffiltext{9}{Dipartimento di Astronomia, Universit\'a di Bologna, via Ranzani 1, I-40127, Bologna, Italy}
\altaffiltext{10}{Centre de Physique Theorique, Marseille, Marseille, France}
\altaffiltext{11}{Institut d'Astrophysique de Paris, UMR 7095 CNRS, Universit\'e Pierre et Marie Curie, 98 bis Boulevard Arago, F-75014 
Paris, France.}
\altaffiltext{12}{Dipartimento di Astronomia, Universita di Padova, Padova, Italy}
\altaffiltext{13}{INAF, Osservatorio di Roma, Monteporzio Catone (RM), Italy}
\altaffiltext{14}{California Institute of Technology, Pasadena, CA, USA.}
\altaffiltext{15}{Harvard-Smithsonian Center for Astrophysics, 60 Garden Street, Cambridge, MA, 02138}
\altaffiltext{16}{Universitats-Sternwarte, Scheinerstrasse 1, D-81679 Muenchen, Germany}
\altaffiltext{17}{Department of Physics, Carnegie Mellon University, 5000 Forbes Avenue, Pittsburgh, PA 15213}
\altaffiltext{18}{Max-Planck-Institut f\"ur Astronomie, K\"onigstuhl 17, D-69117 Heidelberg, Germany}


\begin{abstract}

The impact of environment on AGN activity up to $z\sim1$ is assessed
by utilizing a mass-selected sample of galaxies from the 10k catalog
of the zCOSMOS spectroscopic redshift survey.  We identify 147 AGN by
their X-ray emission as detected by $XMM$-Newton from a parent sample
of 7234 galaxies.  We measure the fraction of galaxies with stellar
mass $M_*>2.5\times10^{10}$ M$_{\sun}$ that host an AGN as a function
of local overdensity using the 5th, 10th and 20th nearest neighbors
that cover a range of physical scales ($\sim1-4$ Mpc).  Overall, we
find that AGNs prefer to reside in environments equivalent to massive
galaxies with substantial levels of star formation.  Specifically,
AGNs with host masses between $0.25-1\times10^{11}$ M$_{\sun}$ span
the full range of environments (i.e., field-to-group) exhibited by
galaxies of the same mass and rest-frame color or specific star
formation rate.  Host galaxies having $M_*>10^{11}$ M$_{\sun}$ clearly
illustrate the association with star formation since they are
predominantly bluer than the underlying galaxy population and exhibit
a preference for lower density regions analogous to SDSS studies of
narrow-line AGN.  To probe the environment on smaller physical scales,
we determine the fraction of galaxies ($M_*>2.5\times10^{10}$
M$_{\sun}$) hosting AGNs inside optically-selected groups, and find no
significant difference with field galaxies.  We interpret our results
as evidence that AGN activity requires a sufficient fuel supply; the
probability of a massive galaxy to have retained some sufficient
amount of gas, as evidence by its ongoing star formation, is higher in
underdense regions where disruptive processes (i.e., galaxy
harrassment, tidal stripping) are lessened.

\end{abstract}





\section{Introduction}

The local environment of galaxies harboring Active Galactic Nuclei
(AGN) and QSOs has long been thought to play a potential role in
triggering mass accretion onto Supermassive Black Holes (SMBHs).  With
many of the properties of galaxies (e.g., morphology, color, star
formation rate) clearly dependent on environment \citep[e.g.,
][]{ka04,ba06,co06,cu09} and the possibility of a common history of
mass assembly for SMBHs and their host bulges
\citep[e.g.,][]{gr04,bo06,ho08,so08}, we expect that AGNs may prefer
to reside in specific environments most nuturing for their growth.
Identifying environmental factors might allow us to determine the
physical mechanism(s) responsible for driving accretion such as major
mergers of galaxies that has been demonstrated through numerical
simulations to be able to remove angular momentum from
rotationally-supported gas thus transfering mass to the nuclear
regions \citep[e.g.][]{ba96,mi96}.  For example, there is evidence
that AGNs reside in dark matter halos with masses
$M_{halo}\sim10^{12-13} M_{\sun}$ \citep{por04,hop07,ma08,pa08,bon08},
a mass regime comparable to the group-scale environments thought to be
fertile ground for galaxies to coalesce.  As well, high density
regions such as massive clusters of galaxies are expected to be
inhospital environments for AGN \citep{dr85} given the strong
empirical association between AGN activity and concurrent star
formation \citep[e.g., ][]{ka03,jahnke04,si09} although
counterevidence exists for the radio-loud population
\citep{hill91,be07}.

Environmental studies to date have presented seemingly disparate
results most likely due to varying selection methods and physical
scales used to characterize environment.  Using a large sample of
narrow-line AGN from the SDSS, \citet{mi03} find that there is no
environmental dependence on AGN activity using a magnitude-limited
sample of galaxies and meaning environment on scales of around a few
megaparsecs.  Upon further investigation, \citet{ka04} find that an
environmental dependence, similar to star formation, emerges when
implementing a selection based on stellar mass and considering the
luminosity of the AGN.  Supportive of this scenario, \citet{coil07}
find that quasars in the DEEP2 survey fields have environments similar
to blue galaxies.  On the other hand, luminous quasars from the SDSS
have an overabundance of galaxies within their vicinity on smaller
scales \citep[$<0.1$ Mpc;][]{se06} that is in agreement with
clustering analysis based on quasar pairs \citep{he06} but may still
have significant biases in the employed methods \citep{pad08}.
Furthermore, the enhancement in star formation attributed to galaxy
mergers on scales of 0.01-0.1 kpc is not reflected in AGN accretion
\citep{li08} for which the authors conclude may be due to varying
timescales with star formation preceeding AGN activity
\citep[e.g.,][]{sch07} occurring within a more relaxed host galaxy.

X-ray selected surveys with both $Chandra$ and XMM-$Newton$ now enable
the study of the environment of AGNs including the obscured population
at higher redshifts ($z\gtrsim0.3$) where optical selection of
narrow-line AGN is difficult.  \citet{gr05} first looked into the
environments of X-ray selected sources detected in the $Chandra$ Deep
Fields and found that the near-neighbor counts were identical to the
galaxies without AGN.  This led the authors to conclude that mergers
were not the physical mechanism triggering mass accretion especially
since the AGN host galaxies were no more asymmetric than the average
galaxy of equivalent luminosity \citep[see][for a morphological study
of AGN hosts in COSMOS]{ga08}.  While \citet{ge07} claim based on a
large spectroscopic sample of galaxies from DEEP2 that X-ray selected
AGN at $z\sim1$ prefer higher density environments, their results are
consistent with those of galaxies having similar host properties
(i.e., absolute magnitude, rest-frame color).  From a complementary
perspective, \citet{ma07} has measured the AGN content of galaxy
clusters and found no significant difference with the fraction of
field galaxies hosting AGN.  Recently, \citet{gi08} find clustering
lengths of AGN in the COSMOS field comparable to massive galaxies and
conclude that the clustering signal is reflective of SMBHs preferring
to reside in galaxies with $M_*>3\times10^{10}$ M$_{\sun}$.  To date,
an environmental analysis of X-ray selected AGN based on both a large
spectroscopic survey of galaxies and careful consideration of
selection to disentangle the degeneracy between mass and environment
seen in galaxy studies \citep[e.g.,][]{ba06,van08,cu09} has not yet
been attempted.

Here, we utilize the rich multiwavelength observations of the COSMOS
field \citep{sc07} to determine the role of environment in triggering
AGN at $0.1<z<1.0$.  The COSMOS survey is roughly a 2 square degree
region of the sky selected to have full coverage with all major
observatories both from the ground (i.e., Subaru, VLT) and space
(e.g., $HST$, $Spitzer$, XMM-$Newton$).  The zCOSMOS survey
\citep{lilly07,lilly09} targets objects for optical spectroscopy with
the VLT in two separate observing programs. A 'bright' sample
($i_{ACS}<22.5$) is observed with a red grism to provide a wavelength
coverage of $5500-9500$ ${\rm \AA}$ ideal to identify galaxies ($L_*$)
up to $z \sim 1.2$.  A deeper program, not utilized in the present
study, targets faint galaxies ($B<25$), selected to be in the redshift
range $1.5\lesssim z\lesssim 2.5$ using a blue grism for an effective
wavelength coverage of $3600<\lambda<6700~{\rm \AA}$.  For the present
study, we select galaxies based on reliable spectroscopic redshifts
from the zCOSMOS 'bright' program and their stellar mass estimates
based on broad-band photometry.  Those that host AGN are identified by
their X-ray emission as detected by XMM-$Newton$
\citep{ha07,cap07,cap08}.  We use the nearest neigbor approach to
determine the projected local galaxy density using the
three-dimensional galaxy distribution as fully described in
\citet{ko09a}.  In addition, the zCOSMOS optically-selected group
catalog \citep{kn09} offers a complementary perspective on the role of
environment.  The low optical luminosity and obscured AGN further
enable us to determine if the host galaxies of AGN exhibit a trend
similar to the color-density or SFR-density relations found for
non-active galaxies.  Finally, we refer the reader to \citet{si09}
that presents the properties (i.e., stellar mass, SFR, rest-frame
color) of the hosts of X-ray selected AGN in the COSMOS field.

Throughout this work, we assume $H_0=70$ km s$^{-1}$ Mpc$^{-1}$,
$\Omega_{\Lambda}=0.75$, $\Omega_{\rm{M}}=0.25$ and use AB
magnitudes.

\section{Data and derived properties}

\subsection{Parent galaxy sample}

\begin{figure}
\hskip -0.7cm
\includegraphics[angle=0,scale=0.53]{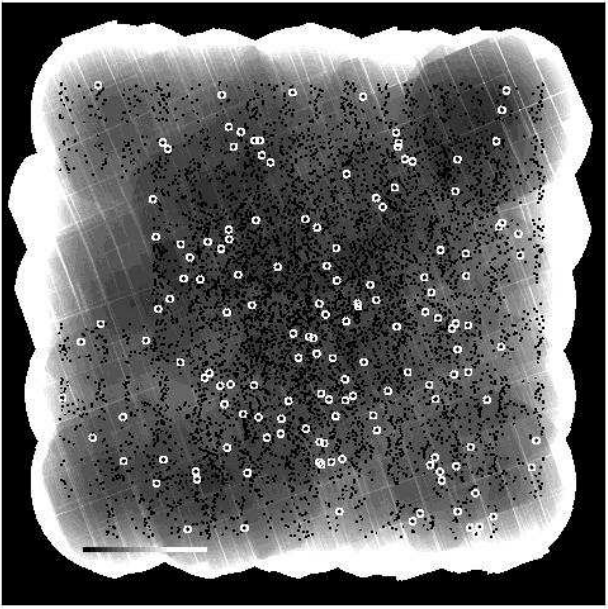}


\caption{Spatial distribution of 7234 zCOSMOS galaxies (small black
dots) with $i_{acs}<22.5$ and $0.1<z<1.0$. Those identified as AGN
based on XMM-$Newton$ detections (147; $L_{0.5-10~{\rm keV}}>10^{42}$
erg s$^{-1}$) are highlighted with a larger white circle.  The
underlying grey scale image is the exposure map of the XMM-$Newton$
mosaic (north is up and east is to the left).  A scale bar
(20$\arcmin$ in length) ranges from $1\times10^{-15}-1\times10^{-14}$
erg cm$^{-2}$ s$^{-1}$.}

\label{selection}
\end{figure}

In order to compare the environments of AGN hosts with those of normal
galaxies, we use for the latter the well-defined "parent sample" of
galaxies up to $z\sim1$ from the zCOSMOS 10k spectroscopic 'bright'
catalog, as done in \citet{si09}.  Specifically, we select 7234
galaxies with an apparent magnitude $i_{ACS}<22.5$, and a redshift
$0.1<z<1.0$ having a spectroscopic redshift quality flag greater than
or equal to 2.0 \citep[see][]{lilly07} that amounts to a confidence of
$\sim99\%$ for the overall sample.  The spatial sampling as shown in
Figure~\ref{selection} is fairly uniform across the central one square
degree while the completed zCOSMOS 20k catalog will fill in the gaps
mainly located along the perimeter of the COSMOS field.  Further
details on the spatial sampling and quality assurance can be found in
\citet{lilly07,lilly09} as well as a complete description of target
assignments, data acquisition and the subsequent reduction procedure
based on the VIMOS Interactive Pipeline and Graphical Interface
package \citep[VIPGI;][]{sco05}.

\begin{figure}
\epsscale{2.2}
\plottwo{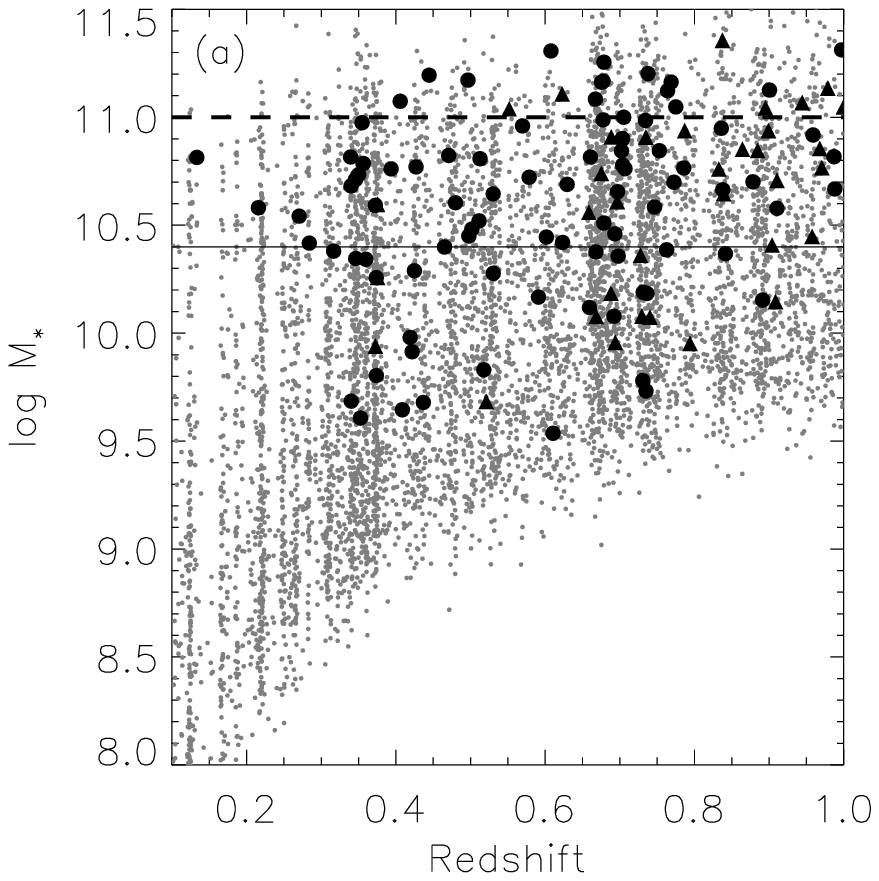}{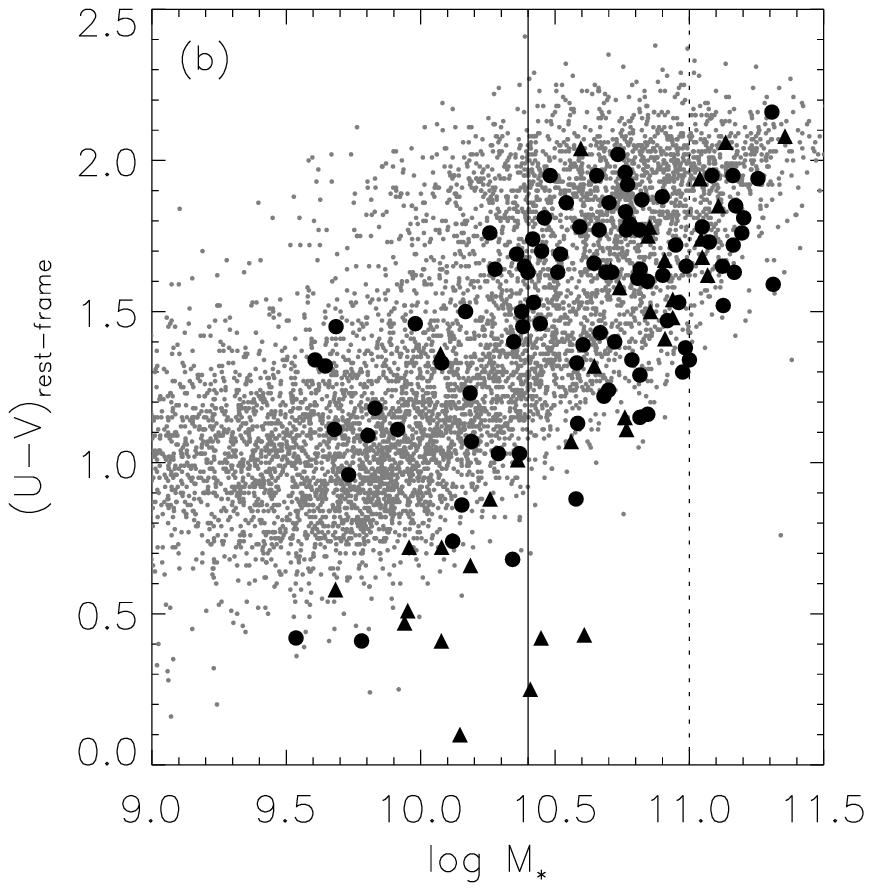}

\caption{(a) Stellar mass versus redshift for 7234 zCOSMOS galaxies
($0.1<z<1.0$, small grey circles).  (b) Rest-frame color $U-V$ versus
stellar mass.  The mass limits for two samples are highlighted in both
panels: $log~M_*>10.4$ (solid line) and $log~M_*>11$ (dashed line).
Galaxies hosting X-ray selected AGN (147) are marked by a larger black
symbol with respect to their X-ray luminosity (circles:
$42<log~L_{0.5-10~{\rm keV}}<43.7$; triangles: $log~L_{0.5-10~{\rm
keV}}>43.7$) in both panels.}

\label{mass_select}
\end{figure}

Stellar masses, including rest-frame absolute magnitudes ($M_U$,
$M_V$) in the AB system, are derived from fitting stellar population
synthesis models from the library of \citet{bc03} to both the
broad-band optical \citep[CFHT: $u$, $i$, $K_s$; Subaru: $B$, $V$,
$g$, $r$, $i$, $z$;][]{capak07} and near-infrared
\citep[$Spitzer$/IRAC: $3.6 \mu $, $4.5 \mu $;][]{sand07} photometry
using a chi-square minimization for each galaxy.  The measurement of
stellar mass (M$_{*}$) includes (1) the assumption of a Chabrier
initial mass function\footnote{Masses are not corrected to those based
on a Salpeter IMF as done in \citet{si09}.}, (2) a star formation
history with both a constant rate and an additional exponentially
declining component covering a range of time scales ($0.1 < \tau < 30$
Gyr), (3) extinction ($0<A_V<3$) following \citet{ca00}, and (4) solar
metallicities.  In Figure~\ref{mass_select}a, we show the distribution
of stellar-mass versus redshift for our sample.  Further details on
mass measurements of zCOSMOS galaxies can be found in \citet{bo09},
\citet{me09} and \citet{po09}.

To minimize any selection biases, we determine a minimum mass
threshold that all galaxies must satisfy.  The mass limit is set to
ensure a fairly complete representation of both blue and red galaxies
at all redshifts considered.  \citet{me09} estimate based on a series
of mock catalogs from the Millennium simulation that the zCOSMOS
'Bright sample' is essentially complete for galaxies with
$log~M_*\approx10.6$ at $z=0.8$ while the completeness drops to
$\sim50\%$ at $z=1$.  We impose a slightly lower mass limit of
$log~M_*>10.4$ (units of M$_{\sun}$) to provide a fair representation of
galaxies covering the full range of rest-frame color $U-V$ up to
$z\sim1$ and ensure an adequate sample that host AGN (see below).  In
Figure~\ref{mass_select}b, it is evident that this mass limit is
essentially imposed by the red galaxy population due to the initial
selection on apparent magnitude.  Above this mass limit, we have a
sample of 2457 galaxies, over the redshift range $0.1<z<1.0$
(Table~\ref{density-sample}; Sample A), suitable for subsequent
analysis.

\begin{deluxetable*}{llllll}
\tabletypesize{\small}
\tablecaption{Sample statistics-density field analysis\label{density-sample}}
\tablewidth{0pt}
\tablehead{\colhead{Sample}&\colhead{Mass\tablenotemark{a}}&\colhead{Redshift}&\colhead{\# Galaxies}&\colhead{\# AGN}&\colhead{Use}\\
&\colhead{range}&\colhead{range}&&($L_{0.5-10~{\rm keV}}$)}
\startdata
All&-----&$0.1<z<1.0$&7234&147 ($>42$)&Total sample\\
A&$>10.4$&$0.1<z<1.0$&2457&63 ($42.48-43.7$); 88 ($>42.48$)&AGN fraction\\
B&$10.4-11$&$0.1<z<1.0$&1971&48 ($42.48-43.7$)&AGN fraction\\
C&$>11$&$0.1<z<1.0$&486&14 ($42.48-43.7$); 20 ($>42.48$)&AGN fraction\\
D&$>10.2$&$0.1<z<0.8$&2482&77 (42-43.7)&Color-density relation\\

\enddata
\tablenotetext{a}{units of log M$_{\sun}$}
\end{deluxetable*}

Further spectroscopic measurements such as emission and absorption
line strengths, and continuum indices are performed through an
automated pipeline \citep["platefit\_vimos";][]{la08} similar to that
performed with the SDSS \citep{tr04}.  For the present study, we
specifically use the [OII]$\lambda$3727 emission-line luminosity,
corrected for slit loss and an AGN contribution if present, to
determine the mass-weighted (specific) star formation rate (sSFR)
using the empirical relation given in \citet{mo06} thus allowing us to
investigate the environmental dependence of star formation for
galaxies hosting AGN.  We refer the reader to \citet{maier09} and
\citet{si09} for a more details regarding spectral measurements of
zCOSMOS galaxies and the removal of AGN emission based on the observed
or inferred [OIII]$\lambda$5007 line flux.

\subsection{AGN identification}

The $XMM$-observations of the COSMOS field \citep{ha07} enable us to
identify those galaxies from the aforementioned zCOSMOS parent sample
of mass-selected galaxies that harbor AGN.  Briefly, a zCOSMOS galaxy
has associated X-ray emission if a maximum likelihood routine
\citep{br07} provides a clear association to an X-ray detection in
either the soft-band ($f_{0.5-2~{\rm keV}}>5\times 10^{-16}$ erg
cm$^{-2}$ s$^{-1}$) or hard-band ($f_{2-10~{\rm keV}}>2\times
10^{-15}$ erg cm$^{-2}$ s$^{-1}$) catalogs \citep{cap08}.  Out of 7234
galaxies, we identify 153 as having significant X-ray emission.  We
note that the frequency of X-ray sources with optical spectra is
higher than would be seen in the general population because a subset
of these sources were designated as 'compulsory' during the design of
VIMOS masks and therefore observed at about twice the sampling rate of
the random targets.  This "bias" does not depend on any other property
of the galaxy.  We account for this when necessary such as measuring
the fraction of galaxies hosting AGN.  We refer the reader to
\citet{br07} and \citet{si09} for more explicite details regarding
X-ray detection, optical source matching and followup spectroscopy.

Our criteria for determining whether X-ray emission is due to an AGN
is based on its point-like spatial extent and whether the luminosity
is above $10^{42}$ erg s$^{-1}$ in either X-ray energy band.  Most
X-ray sources in our sample have luminosities above this threshold as
evident in Figure 1 of \citet{si09} that shows the equivalent data
set.  Therefore, we attribute 147 of the 153 galaxies with X-ray
emission to be the result of an AGN and mark their spatial
distribution in Figure~\ref{selection}.  Our X-ray selected AGNs
mainly have $42\lesssim log~L_X \lesssim 44$ with a few being more
luminous (i.e., QSOs).  Since the majority of these AGN are optically
underluminous, we can investigate if the properties of their host
galaxies (i.e., mass, rest-frame color, SFR) depend on environment
without too much concern for AGN contamination especially for those
with $log~L_{0.5-10 {\rm~keV}}<43.7$.  This upper limit is based on an
empirical relation \citep{si05} where optical emission, associated
with X-ray selected AGN at $\log~\nu l_{\nu}<43.3$ at $E = 2$~keV is
primarily due to their host galaxy since there is a strong departure
from the known $l_{\rm opt}-l_{\rm X}$ relation for more luminous AGN.
We convert this monochromatic luminosity to a value of $10^{43.7}$ erg
s$^{-1}$ in the broad band (0.5--10.0 keV) assuming a power-law
spectrum with photon index $\Gamma=1.9$.  The lack of broad emission
lines for most of our sample \citep[see Fig. 1 of ][]{si09} within
this restricted luminosity range also lends support that these objects
are dominated in the optical by their host galaxy. Here, we further
highlight in Figure~\ref{mass_select} the galaxies that harbor X-ray
selected AGN with respect to the stellar mass\footnote{It is worth
mentioning, as done in \citet{si09}, that there may be a potential
problem that galaxies with even moderate-luminosity AGN may have
inaccurate mass estimates; a bluer continuum will essentially reduce
the stellar age and hence lower the mass measurements since the
derived mass-to-light ratio depends strongly on the spectrum.} and
rest-frame color.  We note that the overdensity of AGNs at $z\sim0.7$
(Fig.~\ref{mass_select}$a$) is indicative of an underlying number
density of galaxies and not an enhancement of AGN activity in
large-scale structures \citep{si08a} although a full assessment of
such effects will be explored in a future study having improved
statistics.

We remark that a significant sample of additional AGN have been
spectroscopically-identified through the Magellan/IMACS observing
program \citep{tr07} that can effectively improve upon the sample.
For the current quantitative analysis, we choose to use only zCOSMOS
identifications in order to maintain a high degree of uniformity and
well-defined relation to a parent sample of galaxies.  We will explore
the feasibility of incorporating such samples in a future
investigation.

\begin{figure*} 

\hskip 0.0cm
\includegraphics[angle=0,scale=0.41]{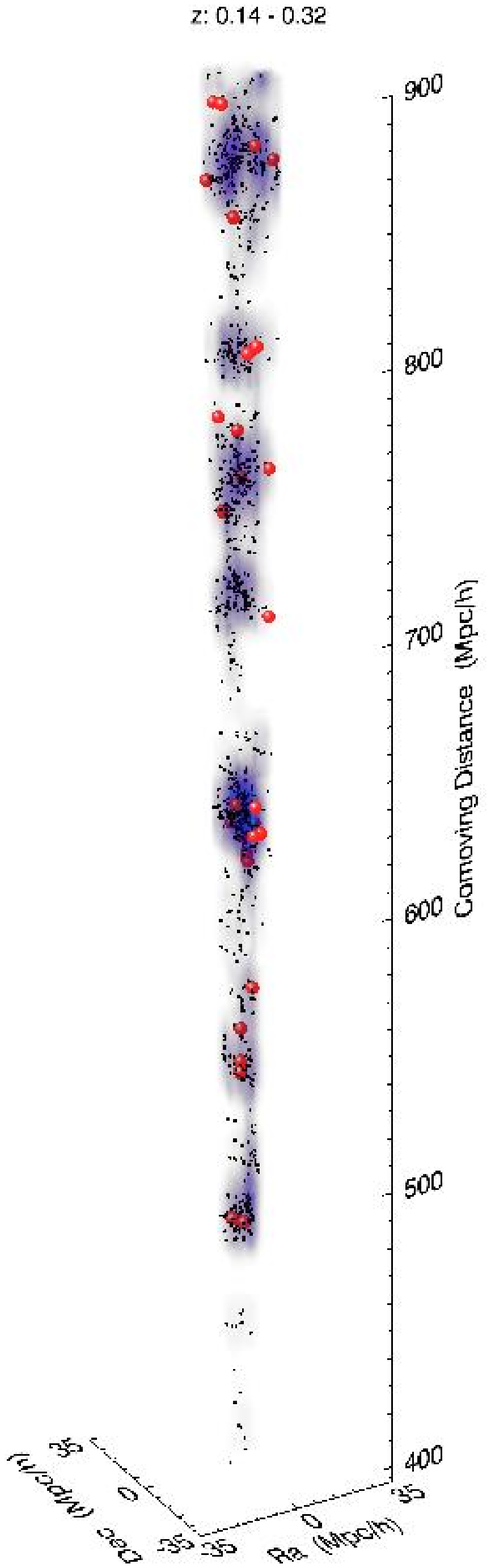}
\hskip -1.3cm
\includegraphics[angle=0,scale=0.41]{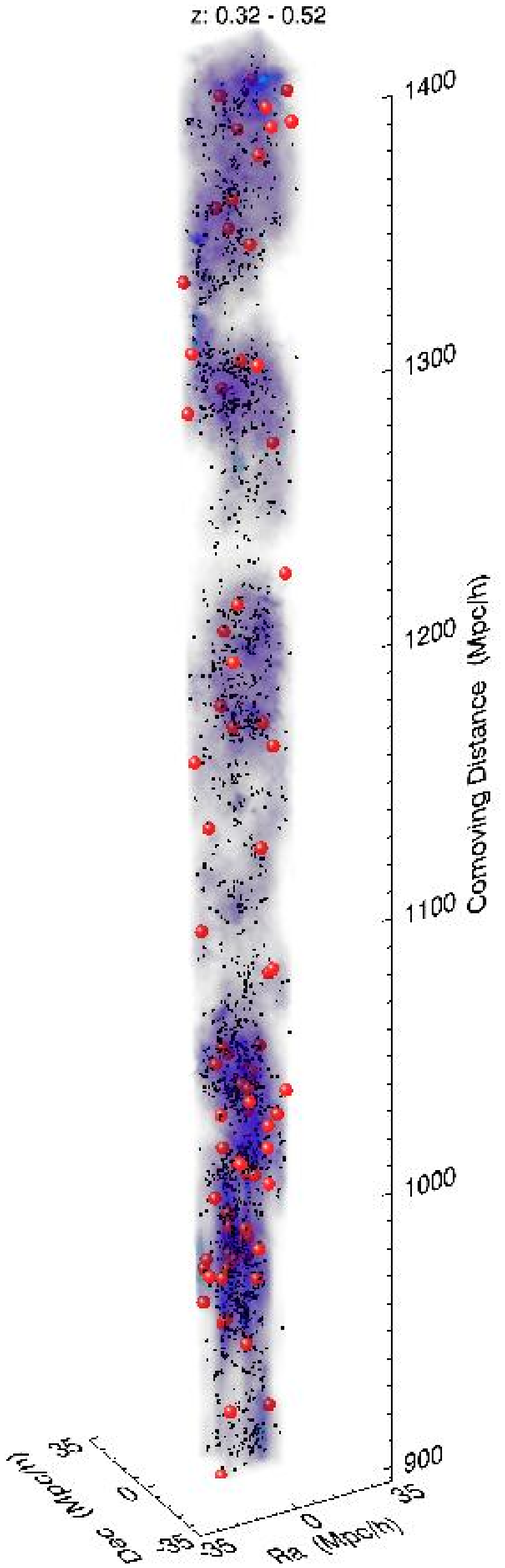}
\hskip -1.2cm
\includegraphics[angle=0,scale=0.41]{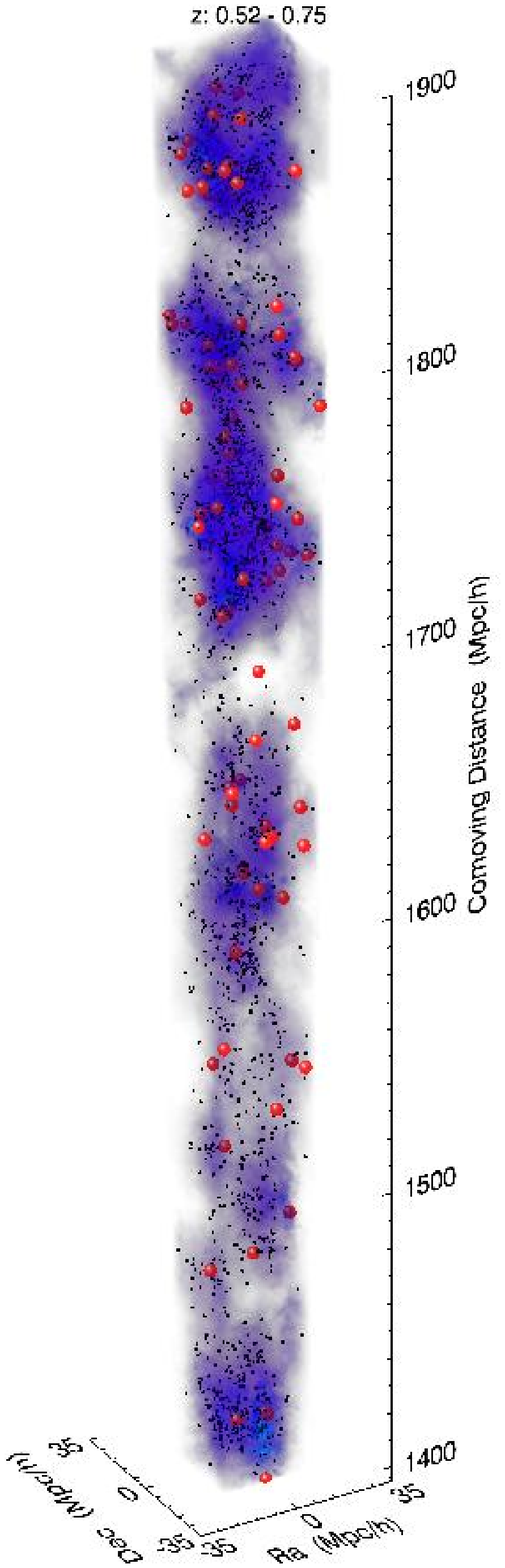}
\hskip -1.2cm
\includegraphics[angle=0,scale=0.41]{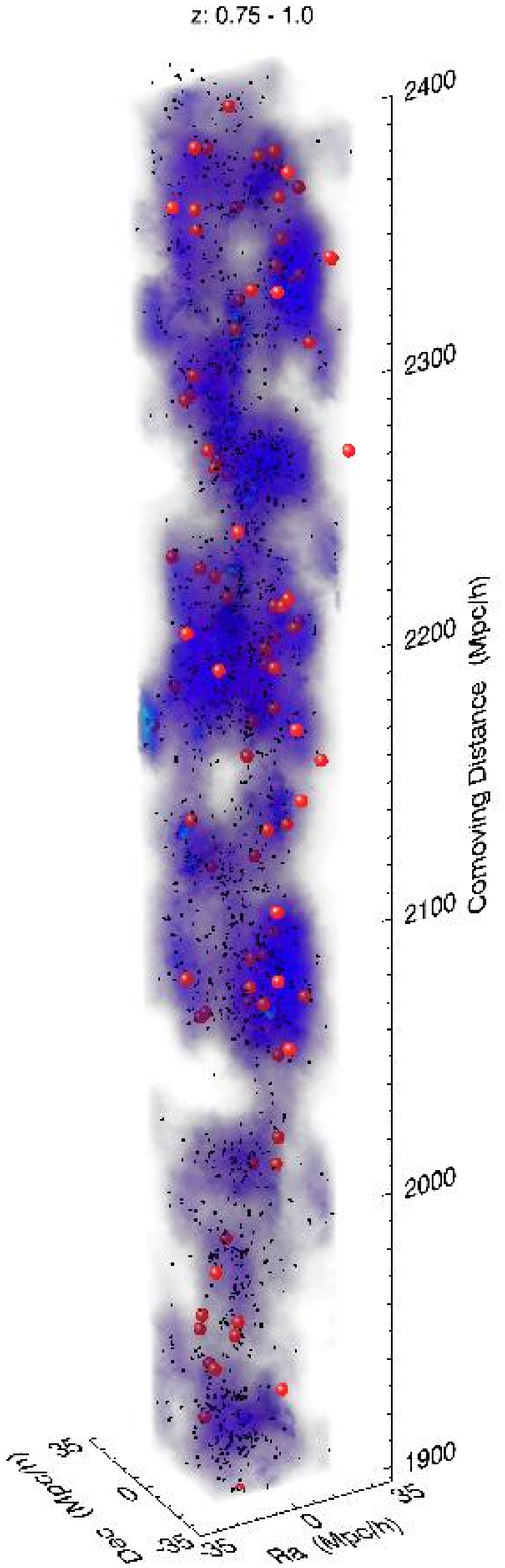}

\caption{Three-dimensional distribution of AGN with respect to the
zCOSMOS overdensity distribution.  Each panel shows a different
redshift interval as labelled.  X-ray selected AGN, identified by both
the zCOSMOS and Magellan/IMACS \citep{tr07} programs in the COSMOS
field, are marked in red.}

\label{3D}

\end{figure*}

\subsection{zCOSMOS galaxy density field}

A major aim of the zCOSMOS survey is to reconstruct the
three-dimensional density field using the 'bright' sample to
characterize the environment up to $z\sim1$ and discern its impact on
galaxy evolution.  Here, we briefly outline the methodology and refer
the reader to \citet{ko09a} for full details regarding the galaxy
density estimates in the zCOSMOS field.  The procedure is based on the
algorithm 'Zurich Adaptive Density Estimator' (ZADE; Kova\v{c} et
al. in preparation) that utilizes spectroscopic (10k) and photometric
redshifts (30k) for accurate distances to practically all galaxies
with $i_{ACS}<22.5$.  For galaxies without spectroscopic redshifts,
their photometric redshift likelihood functions are adjusted using
galaxies in the spectroscopic catalog that are closely along the
line-of-sight and within the effective aperture.  This procedure
effectively accounts for the incomplete sampling ($\sim30\%$) of the
current 10k catalog thus improving density estimates throughout the
zCOSMOS volume.  We use density estimates based on a flux-limited
catalog and adaptive apertures (based on nearest neighbors) in order
to maintain a reasonable sample of AGNs over the full redshift range
($0.1<z<1.0$) at the expense of having a redshift dependent smoothing
scale \citep[see Figure 6 of][]{ko09a}.  To minimize redshift-space
distortions induced by the intrinsic velocity dispersion of groups and
clusters, a projection of $\pm 1000$ km s$^{-1}$ in redshift space is
implemented.  Finally, we express the quantitative measure of the
environment as an overdensity ($\delta$; where $1+\delta=
\rho~$$<$$\rho (z) $$>$$^{-1}$) at the position of each galaxy
relative to the mean density ($<$$\rho (z)$$>$) at a given redshift.
Error estimates are obtained by comparison with reconstructed
overdensities from mock catalogs \citep{ki07} extracted from the
Millenium simulation \citep{sp05} and tailored to match the current
zCOSMOS sample.  Typical errors on $log~(1+\delta)$ are between
0.1-0.15 over a wide range of overdensity \citep[see Figure 4
of][]{ko09a}.

It is important to consider if any difference in spatial sampling
exists between the AGN and randomly-selected galaxies.  Such an effect
may be inherently induced by the VIMOS automated slit assignment
software when designating sources as 'Compulsory', as done for a
subset of the X-ray sources.  We do find that the number of randomly
observed galaxies in the vicinity of 'compulsory' targets is lower
than that of the random sample.  This is slightly noticeable in
Figure~\ref{selection} where the surface density of AGNs is not as
concentrated towards the center of the field as the galaxies.  This
effect is both small and compensated by the 'ZADE' approach thus we do
not expect any significant bias in our density estimates.

In Figure~\ref{3D}, we show the zCOSMOS density field in comoving
coordinates with the location of galaxies and those hosting AGN as
marked.  For display purposes, the overdensity distribution, estimated
on a grid ($\Delta \alpha=\Delta \delta$=1 arcmin, $\Delta z=0.002$)
has been interpolated to a co-moving scale of 1.4 Mpc.  It is clearly
evident that the COSMOS field encompasses a wide range of environments
from voids to dense structures.  The survey area ($\sim2$ deg$^2$;
$\sim80$ comoving Mpc at $z\sim1$) is still narrow enough that sheets,
filaments and walls appear to cut across the field.

\subsection{zCOSMOS galaxy groups}
\label{groups}

A galaxy group catalog \citep{kn09} has been generated using the 10k
'bright' catalog to determine the role of environment in galaxy
evolution \citep[][]{ko09b,io09}.  This approach is complementary to
that based on the density field due to its ability to probe smaller
physical scales.  Two group finding algorithms (friends-of-friends,
Voronoi-Delaunay-Method) are employed to minimize effects such as
group fragmentation, over-merging, spurious detections or missed
groups.  Extensive testing on the aforementioned mock catalogs is done
for optimization of the algorithmns.  For each group, its properties
(velocity dispersion, dynamical mass, corrected richness) are
determined with a level of uncertainty $\sim20\%$ due to the small
number of observed group members ($\sim2-12$) reflective of the bright
magnitude limit of the survey.  A final catalog of 800 groups is
constructed with at least two members using the friends-of-friends
technique well-tuned using the methods employed above.  The
completeness (i.e, detected group members relative to actual members)
has been assessed to be $\sim70-90\%$ \citep[see Fig. 8 of][]{kn09}
and fairly constant with respect to the number of group members.  The
number of interlopers is estimated to be $\sim20\%$.  The redshift
distribution of the final group catalog is very similar to the overall
redshift distribution of zCOSMOS galaxies with two prominent features
at $z\sim0.3$ and $z\sim0.7$.  The masses span an interval
$\sim10^{12}-10^{14}$ $M_{\sun}$ \citep[see Figures 13 and 15
of][]{kn09}, a regime below that of local, massive ($\gtrsim10^{14}$
$M_{\sun}$) clusters.

\section{Large-scale environments of AGN}

\begin{figure*}

\includegraphics[angle=90,scale=0.65]{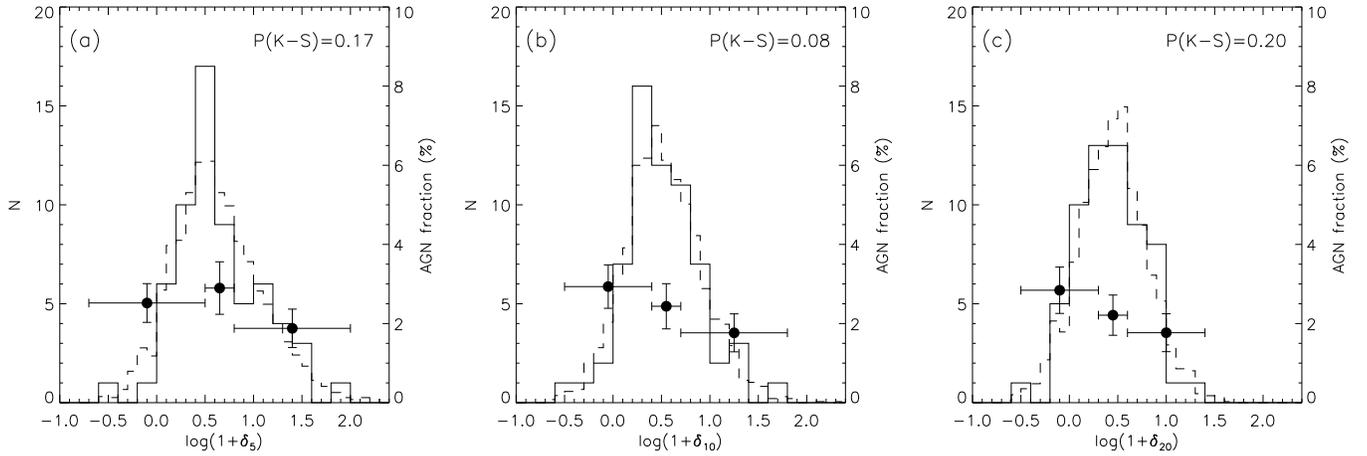}

\caption{Overdensity distribution of a mass-selected sample of 2457
galaxies (Sample A; $log~M_*>10.4$; $0.1<z<1.0$; dashed histogram) and
those that host AGN (63; $42.48<log~L_{0.5-10.0~{\rm keV}}<43.7$;
solid histogram) using the 5th (a), 10th (b) and 20th (c) nearest
neighbors.  In each panel, the underlying parent galaxy population has
been scaled to match the number of AGN.  The results from a K-S test
on the observed distributions are given in each panel.  The data
points are a measure of the fraction of galaxies hosting AGN with
1$\sigma$ errors and horizontal bars denoting the bin size.}

\label{overdensity1}
\end{figure*}

\label{environ}

We present in Figure~\ref{overdensity1} the distribution of zCOSMOS
galaxies and those hosting X-ray selected AGN
(Table~\ref{density-sample}; sample A) as a function of their
environment (i.e., overdensity).  The results are shown using the
distance to the nth nearest neighbor (5th, 10th and 20th) separately
that probe a range of comoving scales ($\sim$1-4 Mpc).  In all panels
(Fig.~\ref{overdensity1} a-c), we generally see that AGN reside in a
broad range of environments (e.g., panel $a$; $0\lesssim log~1+\delta
\lesssim1.5$) remarkably similar to the parent galaxy population of
equivalent stellar mass ($log~M_*>10.4$) as sustantiated by the low
probability ($<20\%$ in each case), based on a K-S test, that we can
reject the null hypothesis (i.e., two distributions are drawn from the
same population).  To avoid any possible luminosity-dependent
systematics, we perform additional K-S tests in fine redshift bins
($\Delta$z=0.1) over the redshift range $0.3<z<0.8$; the probability
(P=0.25, 0.49, 0.54, 0.39, and 0.03) that the two distributions
(overdensities based on the 5th nearest neighbor) are different
conclusively support this finding.

To further investigate if any environment influences are present, we
measure the fraction of galaxies that host AGN as a function of
overdensity.  We follow the technique discussed in $\S$~3.1 of
\citet{le07} to determine the AGN fraction for our parent population
of galaxies.  This method properly accounts for the spatially varying
sensitivity limits of the {\it XMM} observations of the COSMOS field
(Figure~\ref{selection}).  The necessity of this approach is
demonstrated in Figure 1 of \citet{si09} that shows the limiting X-ray
luminosity as a function of redshift for the entire galaxy sample and
the measured X-ray luminosities of those galaxies harboring AGN.  The
sensitivity of the $XMM$ coverage is remarkably uniform as shown by
the relatively narrow distribution of the X-ray limits at each
redshift.  To properly account for the luminosity-redshift relation,
we determine the contribution of each AGN separately to the total
fraction.  The AGN fraction ($f$; see equation{~\ref{fraction} below)
is determined by summing over the full sample of AGN ($N$) with
$N_{\rm gal,i}$ representing the number of galaxies in which we could
have detected an AGN with X-ray luminosity $L^{i}_{\rm X}$.  The
sampling rate of the random galaxies ($S_{gal}$, 29.8\%) and AGN
(i.e., X-ray sources, $S_x$; 71.9\%) are incorporated since these
differ due to the fact that 54\% of the X-ray sources are designated
as 'Compulsory targets' when designing masks for VIMOS.  We estimate
the associated 1$\sigma$ error (equation~\ref{fraction_error}) using
binomial statistics where $N^{eff}_{agn}$ is the number of AGN that
would be detected if all galaxies have the same limiting X-ray
sensitivity and the sample of AGN was randomly selected.  Here, we
only consider AGN with $42.48<log~L_{0.5-10 {\rm keV}}<43.7$.  The
lower limit ensures that we have a statistically significant sample of
parent galaxies ($\gtrsim700$) that could host each AGN while the
upper limit restricts the sample to low-to-moderate luminosities thus
securing the accuracy of their host-galaxy masses.  We refer the
reader to \citet{si08a} for further details and results employing this
method.

\begin{equation}
f=\sum_{i=1}^{N}\frac{1/S_x}{N_{\rm gal,i}/S_{gal}}
\label{fraction}
\end{equation}

\begin{equation}
\sigma^2=N^{eff}_{agn}\times(N_{gal}-N^{eff}_{agn})/N^3_{gal}
\label{fraction_error}
\end{equation}

In Figure~\ref{overdensity1}$a-c$, we give the results of this
exercise with the fraction of galaxies hosting AGN as measured in
three bins of overdensity with widths set to have equal numbers of
galaxies therein.  In agreement with the previous analysis, we find
that the AGN fraction is not strongly enhanced at any particular value
of overdensity.  The AGN fraction is low ($\sim2-4\%$) compared to
similar studies \citep[e.g.,][]{ka04} and due to our selection in a
narrow range of X-ray luminosity.  We finally remark that an
underlying dependence on environment for a subset of our sample is
realized as demonstrated in the next section and possibly seen here by
the slight trend in the AGN fraction (Figure~\ref{overdensity1}$b-c$).


We further test whether an environmental influence is present for AGN
of a specific X-ray luminosity or hardness ratio.  The hardness ratio
(HR) is a means of easily characterizing the X-ray spectrum (i.e.,
level of absorption) by measuring the ratio of X-ray counts in the
hard band (H; 2.0-10 keV) relative to the soft band (S; 0.5-10 keV):
HR= (H-S) / (H+S).  We plot in Figure~\ref{agn_prop} local overdensity
versus X-ray luminosity while highlighting hard ($HR>-0.2$) sources.
We find that there is no correlation between galaxy environment and
the intrinsic AGN emissivity or any absorbing material.

\begin{figure}
\epsscale{1.1}
\plotone{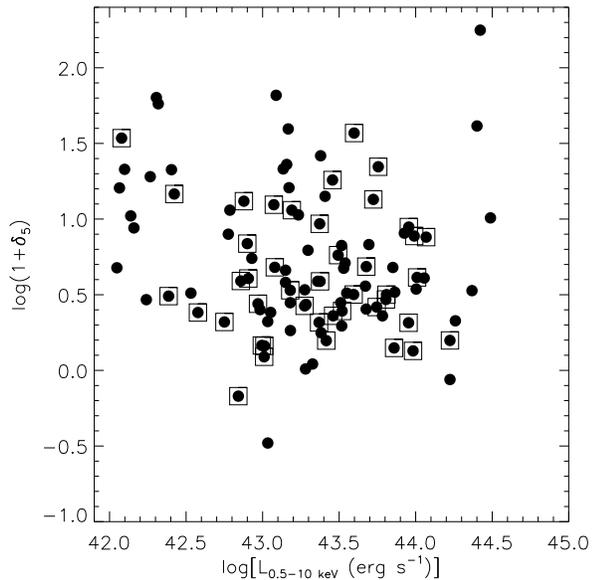}

\caption{Overdensities using the 5th nearest neighbor for galaxies
hosting AGN as a function of their X-ray luminosity.  Absorbed AGN
($HR>-0.2$) are further marked by an open square.  There are no
environmental influences dependent on the intrinsic properties of AGN.}

\label{agn_prop} 

\end{figure}

\subsection{High mass galaxies}

We investigate whether there exists a relation between AGN activity
and environment for galaxies of a specific stellar mass.  Our
motivation is based on the mass-dependent environmental relationship
seen in SDSS studies \citep{ka04}.  In contrast to our analysis in the
previous section, we include more X-ray luminous AGN
($log~L_{0.5-10~{\rm keV}}>42.48$) that have optical spectra dominated
by their host galaxy as determined by visual inspection thus slightly
improving our statistics for the highest mass galaxies.  In
Figure~\ref{overdensity2}, we show plots equivalant to those presented
in the previous section and split into two mass bins (panels $a-c$:
$10.4<log~M_*<11$; panels $d-f$: $log~M_*>11$) with the statistics
given in Table~\ref{density-sample} (Sample B, C).  For the lower mass
interval (Fig.~\ref{overdensity2}a-c), we find that the results are
consistent with our previous findings and the AGN fraction is
remarkably similar over the full range of overdensities.  On the other
hand, the overdensity distribution (Fig.~\ref{overdensity2}$d-f$) of
galaxies hosting AGN is dissimilar to the underlying massive galaxy
population due to the lack of AGNs in higher density environments
$log(1+\delta>0.5$).  The difference in these distributions is
significant at the 2.3-2.5$\sigma$ level based on the probabilities
(0.99, 0.98) from K-S tests using either the 5th or 10th nearest
neighbor.  We note that these results are still apparent when
restricting the analysis to lower luminosity AGN although at a
slightly reduced level of significance (2$\sigma$).  Therefore, the
fraction of galaxies hosting AGN is higher in underdense regions
($log(1+\delta<0.5$) since the offset between the distributions is
substantially greater than typical errors in overdensity.  These
results may further explain the discrepancy between the environmental
studies based on the SDSS \citep{mi03,ka04}.  We remark that our
current sample of massive galaxies ($log~M_*>11$) in zCOSMOS with AGN
is small (20) and will improve with the full zCOSMOS 20k catalog and
deeper $Chandra$ observations.

\begin{figure*}
\includegraphics[angle=90,scale=0.73]{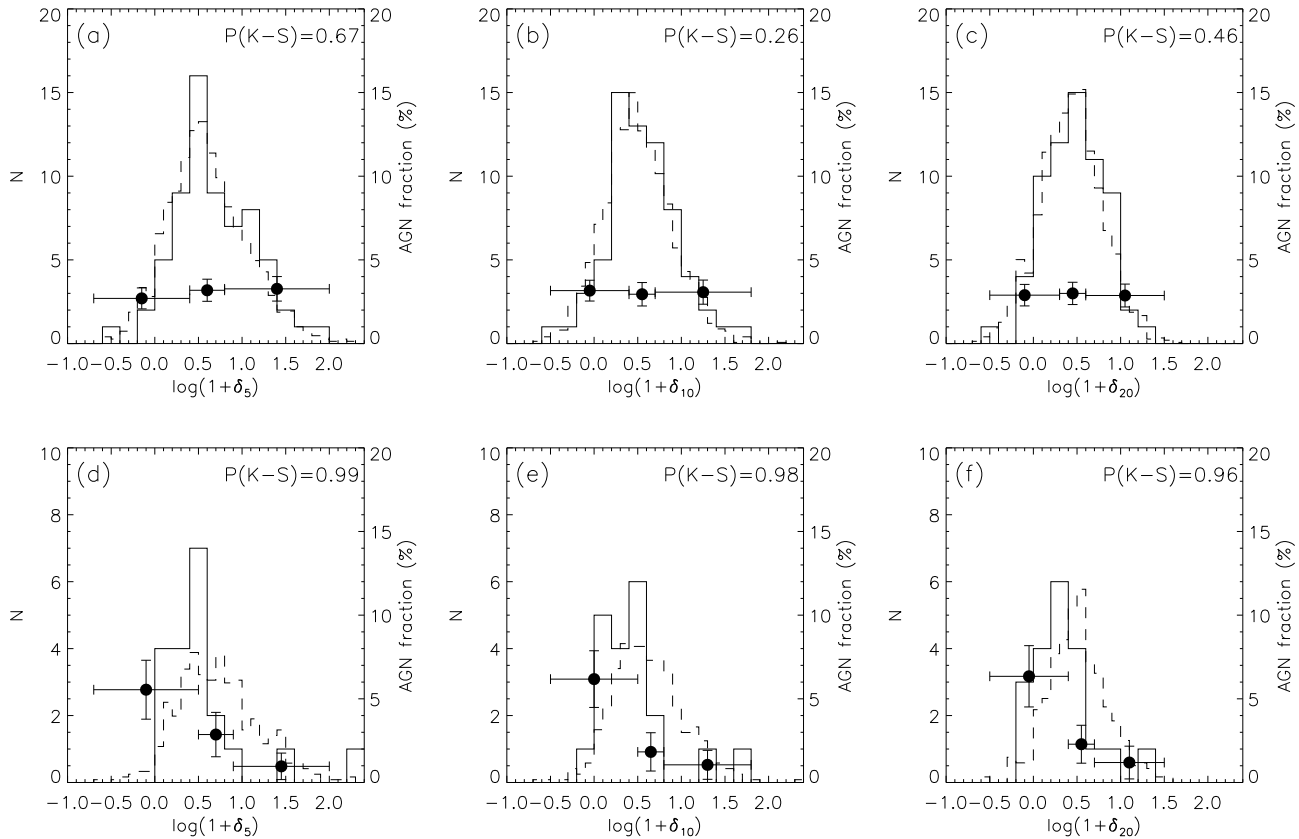}

\caption{Overdensity distribution of galaxies ($0.1<z<1.0$) and those
hosting AGN as a function of stellar mass: (Sample B; panels a-c)
$10.4<log~M_*<11.0$, (Sample C; panels d-f) $log~M_*>11.0$.  The
histograms and data points are the same as given in
Fig.~\ref{overdensity1}.  The results from a K-S test are given in
each panel.}

\label{overdensity2} 

\end{figure*}

\subsection{Dependence on stellar content of AGN hosts}

The result that AGN prefer to reside in lower density environments for
those having massive host galaxies is reminiscent of the well-known
color-density relation for galaxies in general.  To explore this
aspect further, we plot the distribution of rest-frame color $U-V$
versus overdensity in Figure~\ref{color_density}.  For the following
analysis, we decrease our mass limit $log~M_*>10.2$ in order to
improve upon the numbers of AGN hosts with blue colors thus requiring
us to reduce the maximum redshift to $z=0.8$ to not induce a color
bias (Table~\ref{density-sample}; Sample D).  In panel $a$, galaxies
exhibit an environmental influence with red galaxies ($U-V>1.6$)
having a prominent extension of their overdensity distribution towards
higher values $log(1+\delta_5\gtrsim1.0)$.  The well-known trend of
increasing density for redder galaxies (i.e. color-density relation)
is clearly seen in panel $b$ where we measure the mean overdensity for
galaxies in bins of rest-frame color (black points).

\begin{figure*}
\epsscale{0.8}
\plotone{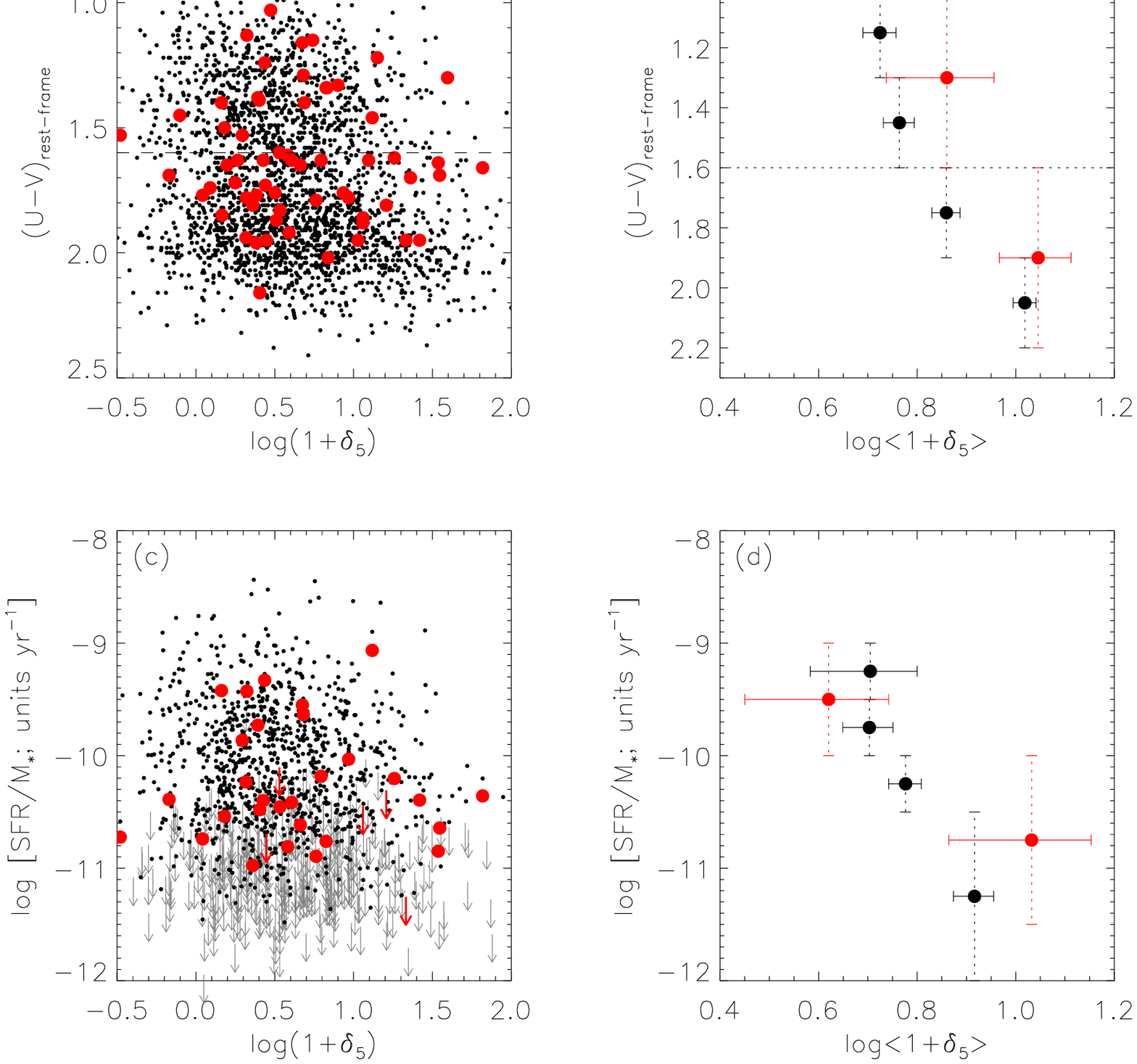}

\caption{($a$) Color-density relation of galaxies (small black
circles; Table~\ref{density-sample}; Sample D) and those with AGN
(large red circles).  ($b$) Mean overdensity and associated error for
color bins marked by the vertical dashed lines.  The symbol colors are
the same as in panel $a$.  The AGN sample is split into two color bins
that essentially represents 'blue cloud' and red sequence galaxies
with a fixed color division ($U-V=1.6$). ($c$) sSFR-density relation
(arrows are upper limits) with mean values given in panel $d$.  We
find that the host galaxies appear to exhibit a color/sSFR - density
relation similar to the underlying galaxy population although improved
statistics are required to firmly justify such a conclusion.}

\label{color_density} 

\end{figure*}

We test whether galaxies hosting AGN exhibit a similar color-density
relation as described above.  AGN hosts are marked appropriately in
Figure~\ref{color_density}.  From their mean overdensities (panel
$b$), AGNs appear to follow a similar relation to the non-active
galaxies although the errors are substantial with a significance of
1.7$\sigma$ in the difference \citep[see][for an equivalent analysis
and conclusion]{ge07}.  In support of the color-density relation for
AGN hosts, we investigate the connection between ongoing star
formation and environment. In Figure~\ref{color_density}$c$, we plot
the mass-weighted 'specific' SFR (sSFR), a quantity based on
[OII]$\lambda$3727 luminosity, versus galaxy overdensity.  The mean
overdensity in bins of sSFR is calculated and shown by the large black
dots in panel $d$.  We find a similar result and uncertainty as above
with the environment being related to the stellar content of the hosts
of AGN.  It is worth highlighting that a clear dichotomy is present
between the importance of both mass and environment in determining the
likelihood of a galaxy's properties \citep[see][for further analysis
based on zCOSMOS galaxies]{cu09}.

Coupled with the results of our companion study \citep{si09}, AGN
prefer to reside in galaxies undergoing star formation irrespective of
their environment.  We illustrate this conclusion in
Figure~\ref{agn_fraction_color} where the fraction of galaxies hosting
AGN is shown as a function of rest-frame color $U-V$ and overdensity.
Galaxies bluer ($U-V\lesssim 1.6$) than those that have stopped
forming stars (i.e., along the red sequence) have higher levels of AGN
activity.  We further find here that this result holds for AGN hosts
residing in either lower or higher density environments.

\begin{figure}
\epsscale{1.0}
\plotone{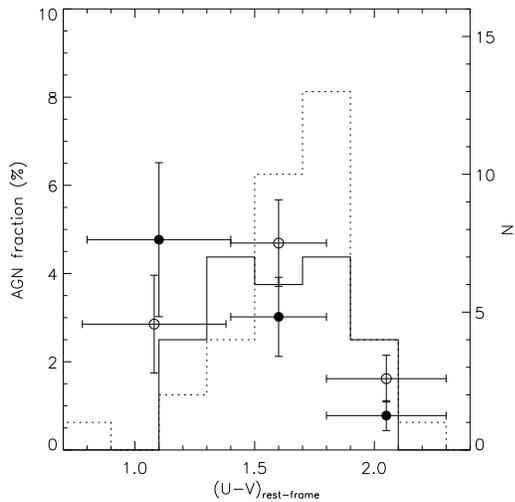}

\caption{Fraction of galaxies hosting AGN as a function of rest-frame
color $U-V$ and environment.  Data points are given for environments
above (filled circle) and below (open circle) $log (1+\delta_5)=0.6$.
The color distribution of AGNs in these two environmental categories
is also shown (solid histogram=high density, dashed histogram=lower
density).}

\label{agn_fraction_color}
\end{figure}.  

By looking in detail at the most massive galaxies ($log~M_*>11$;
Fig.~\ref{color_density_massive}), we can further understand the
relation of AGNs to their environment and the presence of star
formation.  The locus of AGNs in the color-density plane (panel $a$)
is offset from that of massive galaxies due to their bluer colors and
lower overdensities.  Clearly, the distinction between the overdensity
distribution of AGN hosts and their parent population is due to the
fact that AGNs do not reside in red galaxies (panel $b$) that tend to
live in denser environments but rather require fuel for accretion that
tends to drive subsequent star formation.  Therefore, the environments
of AGNs hosts are similar to star-forming galaxies ($c$).  We
interpret these results as evidence that massive galaxies are more
likely to have accreting SMBHs if not subjected to harsher
environments that can effectively expel gas, due to processes related
to galaxy interactions, thus impacting star formation and subsequent
AGN activity.

\begin{figure*}

\includegraphics[angle=90,scale=0.6]{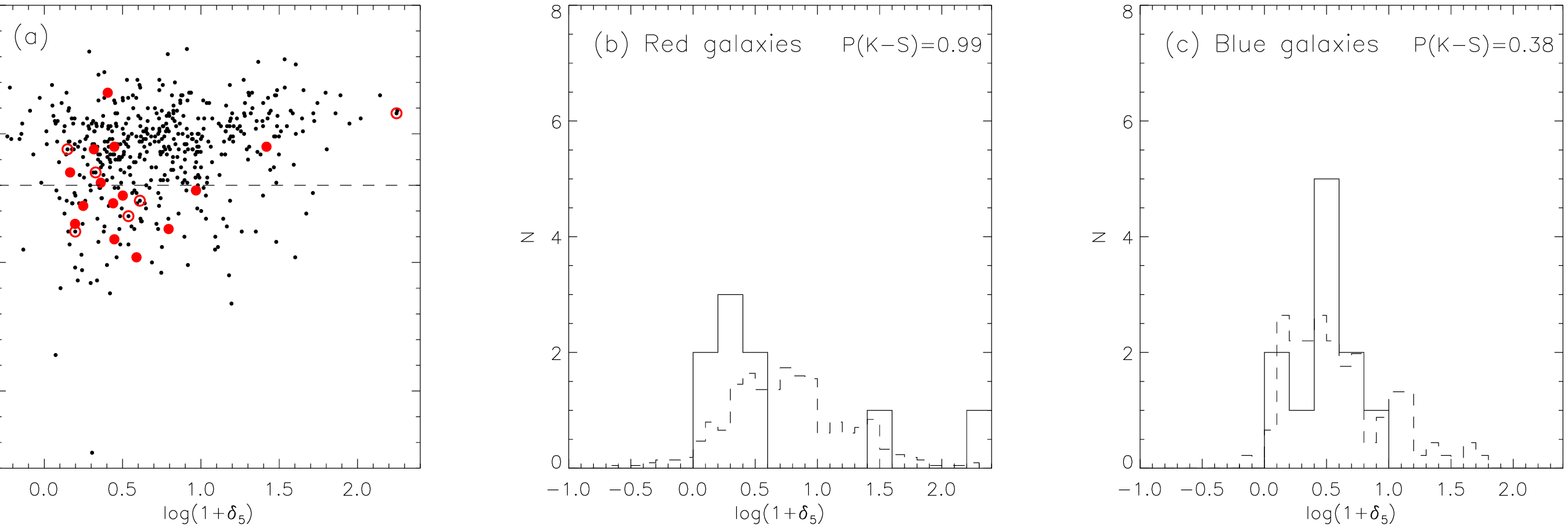}

\caption{Color-density relation for massive galaxies ($M_*>10^{11}$
M$_{\sun}$). ($a$) Equivalent to Figure~\ref{color_density}$a$ with
the addition of more luminous AGN ($log~L_{0.5-10~{\rm keV}}>43.7$)
marked here with an open red circle.  Overdensity distribution of red
($b$; $U-V>1.8$) and blue ($c$; $U-V<1.8$) galaxies (dashed histogram)
separately with those hosting AGN shown by the solid histogram.  The
probabilities based on K-S tests are shown in the respective panel.}

\label{color_density_massive} 

\end{figure*}

\section{AGN content of optically-selected groups}

\begin{figure}
\epsscale{1.0}
\plotone{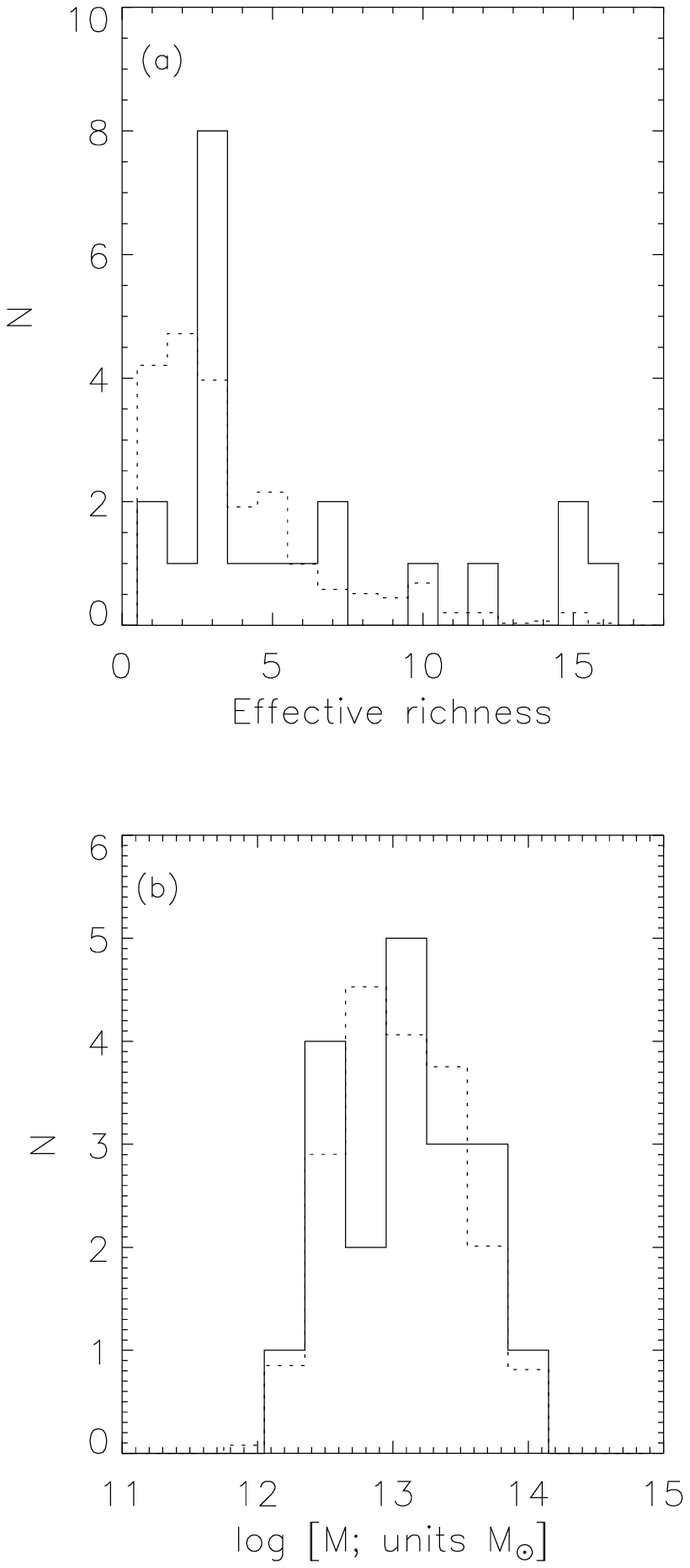}

\caption{Distribution of group properties ($a$: effective richness,
$b$: mass) of galaxies hosting AGN (solid histogram) having
$log~M_*>10.4$.  For comparison, the distribution of all galaxies of
similar masses has been renormalized and shown by the dashed
histogram.}  

\label{group_prop} 

\end{figure}

The presence of AGN within well-defined structures such as groups
\citep{ge08}, clusters \citep{ma07,gi07,koc08,le08}, or large-scale
sheets/filaments \citep{gi03,si08a} as compared to the field offer an
alternative perspective on the environmental impact on AGN activity.
To do so, we utilize the catalog of 800 galaxy groups \citep{kn09} as
described in Section~\ref{groups}.  As done in previous analyses, we
restrict the sample to galaxies with stellar masses above
$2.5\times10^{10}$ M$_{\sun}$ thus resulting in a sample of 2444
galaxies with a designation as either a 'field' or 'group' galaxy. In
Table~\ref{group-sample}, we list the numbers of galaxies in each
subclass, those that host AGNs ($Log~L_{0.5-10~keV}>42.48$) and the
derived AGN fraction as detailed below.

\begin{deluxetable}{llll}
\tabletypesize{\small}
\tablecaption{Sample statistics-group analysis\label{group-sample}}
\tablewidth{0pt}
\tablehead{\colhead{Environment}&\colhead{Galaxies}&\colhead{AGN\tablenotemark{a}}&\colhead{AGN fraction (\%)}\\
&$M_*>10^{10.4}$}\\
\startdata
Field&1675&63&$3.28\pm0.43$\\
Group&828&27&$2.59\pm0.55$\\
Group ($R\tablenotemark{b}\ge1$)&614&21&$2.89\pm0.68$\\
~~~~~~~~~($R\tablenotemark{b}\ge2$)&491&19&$3.17\pm0.79$\\
~~~~~~~~~($R\tablenotemark{b}\ge4$)&237&10&$3.00\pm1.11$\\
\enddata
\tablenotetext{a}{$log~M_*>10.4$; $log~L_{0.5-10~{\rm keV}}>42.48$}
\tablenotetext{b}{Effective group richness}
\end{deluxetable}

We simply measure the fraction of galaxies hosting AGN for the 'group'
and 'field' populations separately to test whether any environmental
dependency exists.  Over the full redshift range ($0.1<z<1.05$), we
measure the fraction to be $3.28\pm0.43\%$ for field galaxies.  For
galaxies in groups, we find an AGN fraction of $2.59\pm0.55\%$, based
on two or more spectroscopically-identified group members, slightly
less than that in the field although not statistically significant.
To account for redshift-dependent effects and spatial sampling in the
group catalog, we further isolate groups having an effective (i.e.,
corrected) richness ($N_{eff}$) defined as the number of galaxies
above a given magnitude threshold \citep[see section 4.2 of][for
details]{kn09} that can be detected over the full redshift range
considered here.  This limit is based on the absolute magnitude in the
$b$-band ($M_{b,lim}<20.5-z$) and includes a redshift term to account
for the luminosity evolution of galaxies.  In
Figure~\ref{group_prop}$a$, we show the distribution of effective
richness for all galaxies ($log~M_*>10.4$) residing in groups that
illustrates the scale of the zCOSMOS groups.  In addition, the
distribution of group mass (Fig.~\ref{group_prop}$b$) is given.
Briefly, the group mass is the mean mass of halos in the mock catalog
that are associated with groups of a specific effective richness
($N_{eff}$).  A redshift dependency is inherent given that effective
richness is based on an observed quantity (i.e., apparent magnitude).
Full details regarding these derived properties can be found in
\citet{kn09}.  Considering a selection on effective richness, we find
a similar fraction of galaxies hosting AGN as compared to the field
with no obvious dependence on the effective richness (see
Table~\ref{group-sample}) of the group.  Furthermore, there is no
significant difference in the X-ray luminosity or host galaxy color
(rest-frame $U-V$) distributions of AGNs in or out of groups
\citep[see][for similar analyses]{ge08}.  These results are in
agreement with our previous findings based on the density field that
AGNs in galaxies with $M_*>10^{10.4}$ M$_{\sun}$ do not show a strong
dependence on environment.

\begin{figure}
\epsscale{1.0}
\plotone{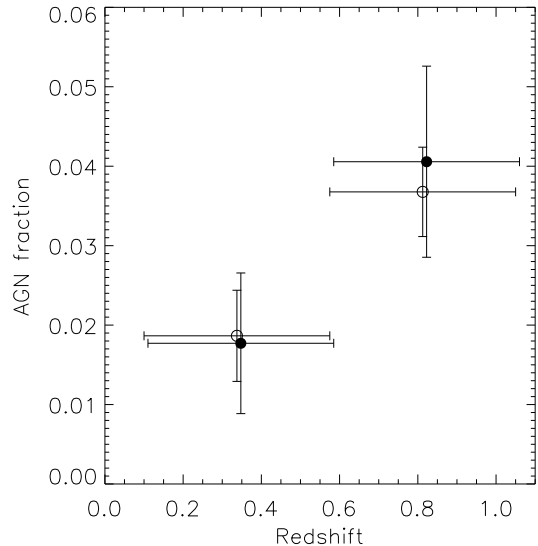}

\caption{Fraction of galaxies ($log~M_*>10.4$) hosting AGN split by
those in galaxy groups ($N_{eff}>2$,filled circle) and those in the
field (open circle).  Errors are $1\sigma$ based on binomial
statistics.}

\label{group_fraction}
\end{figure}

To determine if any signs of evolution exist, we plot the fraction of
group galaxies hosting AGN in two redshift bins
(Figure~\ref{group_fraction}).  Here, we select groups having
$N_{eff}\ge2$ as described above.  The AGN fraction is remarkably
similar between for galaxies in groups and in the field for both
redshift intervals.  The increase in the AGN fraction in the higher
redshift bin for both the field and groups, is most likely due to the
overall positive luminosity evolution of the AGN population
\citep[e.g.,][]{ba05,si08b}.  Therefore, we do not find an enhancement in
the AGN content of groups with redshift as strong as reported for
clusters \citep{ea07} above the level expected by the AGN luminosity
function.  As previously mentioned, we will improve upon the current
sample, as justified by the substantial errors, in a future
investigation by using the zCOSMOS 20k catalog and deeper $Chandra$
observations \citep{elvis09} that provide additional
AGNs of moderate luminosity ($42 \lesssim L_X \lesssim 43$) over the
redshift range 0.5-1.0.  A larger sample will enable us to isolate high
mass galaxies ($M_*>10^{11}$ M$_{\sun}$) as done with the density
field.

\section{Conclusion and summary}

We have assessed whether the environment plays any role in regulating
mass accretion onto SMBHs, on scales of a few megaparsecs, using 7234
galaxies ($i_{acs}<22.5$; $0.1<z<1.0$) from the zCOSMOS survey
('Bright' program).  To do so, a parent sample of 2457 galaxies is
carefully constructed based on stellar mass ($log~M_*>10.4$) with no
obvious color ($U-V$) bias.  Galaxies undergoing substantial accretion
onto a SMBH are identified by the presence of an AGN with X-ray
emission detected by XMM-$Newton$; we find 147 AGN with 101 residing in
host galaxies having $log~M_*>10.4$.  With two characterizations of the
local environment, we measure the fraction of galaxies as a function
of their galaxy overdensity based on the zCOSMOS density field
\citep{ko09a} and presence in optically-selected galaxy groups
\citep{kn09}.

Based on the zCOSMOS density field, we find that the AGNs, with
exception of those residing in the most massive host galaxies
($log~M_*>11$), span a broad range of environments \citep[see][for
similar conclusions]{ge07}, from the field to massive groups,
equivalent to galaxies of similar stellar mass.  Our findings do
appear to be consistent with low redshift studies based on narrow-line
AGN from the SDSS \citep{mi03,ka04,co08} that taken together
demonstrate that the environments depend on the stellar mass of their
host galaxy.  The lack of an environmental dependence on AGN activity
over the lower mass regime ($10.4<log~M_*<11$) in zCOSMOS is
compatible with the results of \citet{mi03} since their
magnitude-limited sample, dominated by low luminosity AGN, most likely
includes many lower mass galaxies.  An environmental dependence with
AGN activity less common in higher density regions emerges for 'strong
AGN' residing in galaxies of higher mass \citep[][]{ka04,co08} and is
also evident in our study based on zCOSMOS galaxies with $log~M_*>11$.
We contend that the discrepant findings based on SDSS studies at low
redshift may be due to the differences in the underlying mass
distribution of galaxes rather than the differences of the AGN
selection.

We find that the presence of young stars in the host galaxies of AGN
\citep[][]{ka03,jahnke04,si09} is also indicative of their environments.
Galaxies hosting AGN in zCOSMOS follow a similar color-density
relation to non-active galaxies \citep{cu09} with blue hosts
($U-V<1.6$) having a lower mean overdensity than red galaxies.  An
equivalent relation based on SFR or that weighted by stellar mass
(specific) supports this assertion.  As fully discussed in
\citet{cu09}, the sSFR is seen to be more strongly dependent
on environment than SFR due to the underlying mass dependency
discussed above.  The majority of massive ($log~M_*>11$) AGN hosts
having colors bluer than the underyling galaxy population agrees with
their residence in lower density environments.  Therefore, we conclude
that the incidence of AGN activity as a function of environment
depends on the properties of its host galaxy namely its mass content
both in the form of stars and gas in the interstellar medium.

Furthermore, we carry out a complementary analysis of the environments
of AGN by utilizing the optically-selected catalog of galaxy groups
\citep{kn09} in zCOSMOS.  Such groups have halo masses
$\sim10^{12}-10^{14}$ M$_{\sun}$ similar to those that host the more
luminous quasars \citep[e.g.,][]{por04,pa08,bon08} and less massive than
clusters at low redshift.  We measure the incidence of AGN activity in
galaxies in groups compared to that in the 'field'.  This method
effectively enables us to probe smaller physical scales ($\lesssim1$
Mpc) not possible with the density field due to the spatial resolution
of $\pm1000$ km s$^{-1}$ along the line-of-sight.  An excess signal
around quasars, based on either near-neighbor counts \citep{se06} or
the quasar correlation function \citep{he06}, highlights the
importance of probing scales $\sim100$ kpc.  Although, the
environments of 31 X-ray selected AGN \citep{wa05} with $0.4<z<0.6$
from the Canada-France-Redshift Survey fields are indistinguishable
from a well-matched control sample of galaxies on scales around 30-500
kpc.

We find in zCOSMOS that AGN are essentially equally likely to reside
in or out of a galaxy group irrespective of its properties (i.e.,
effective richness).  The fraction of group galaxies ($log~M_*>10.4$)
hosting an AGN is similar to 'field' galaxies ($\sim3\%$).  This is in
agreement with both our results based on the density field for
galaxies spanning the full mass range $log~M_*>10.4$, and those of
\citet{ge08} which consider the host galaxy properties.  We note that
the current AGN sample in zCOSMOS residing in galaxy groups is too
small to investigate if an environmental effect for the most massive
galaxies exists as exercised using the density field.  Our lack of a
strong environmental effect is also consistent with the incidence of
X-ray selected AGN in galaxy clusters \citep{ma07} as compared to the
field.  Also, an increase in the AGN fraction in the higher of two
redshift bins is significantly smaller than the enhancement reported
for galaxy clusters \citep{ea07} and in our case most likely due to
the luminosity evolution of the AGN population since both the group
and field samples show an increase of equal magnitude.

We conclude that internal processes \citep{ho_he06} play an important
role in the growth of SMBHs during an AGN phase of moderate-luminosity
($L_X\sim10^{43}$ erg s$^{-1}$; 'Seyfert mode') that accounts for the
bulk of the Cosmic X-ray Background \citep[e.g.,][]{gi07}.  Major
mergers of galaxies, possibly relevant for the more luminous quasar
phenomenon, may not be the primary mechanism for fueling these AGN due
to the lack of any enhancement of activity in specific environments
likely to be conducive for merging (i.e., galaxy overdensities
comparable to the small group scale).  In the opposite sense, we do
find that for the most massive galaxies ($log~M_*>11$) the environment
plays a role possibly through various physical processes (e.g., tidal
stripping, harrassment) in higher density environments that
concurrently shuts down star formation and accretion onto supermassive
black holes.  This highlights the requirements for a galaxy to host an
accreting SMBH, as observed by their X-rays, a massive bulge
\citep[e.g.,][]{sa04,gr05,si08a,ga08} and a sufficient fuel supply as
indicated by the young stars usually present in AGN host galaxies
\citep{ka03,si09}.

\acknowledgements

We thank the referee and Paul Martini for constructive comments that
significantly improved the paper.  This work is fully based on
observations undertaken at the European Southern Observatory (ESO)
Very Large Telescope (VLT) under the Large Program 175.A-0839 (P.I.,
Simon Lilly).

{\it Facilities:} \facility{XMM}, \facility{VLT:Melipal (VIMOS)}

\end{document}